\documentclass[12pt,preprint]{aastex}

\shorttitle{FP in Low-Excitation Radio Galaxies}

\shortauthors{S.-L. Li \& M. Gu}

\begin{document}

\title{Black Hole Fundamental Plane in Low-Excitation Radio Galaxies}

\author{Shuang-Liang Li\altaffilmark{1,2}
and Minfeng Gu\altaffilmark{1}}


\altaffiltext{1}{Key Laboratory for Research in Galaxies and
Cosmology, Shanghai Astronomical Observatory, Chinese Academy of
Sciences, 80 Nandan Road, Shanghai, 200030, China; lisl@shao.ac.cn}
\altaffiltext{2}{University of Chinese Academy of Science, 19A Yuquanlu, Beijing 100049, China}

\begin{abstract}

The radio-X-ray slope in the fundamental plane of radio-loud active galactic nuclei (AGNs) is found to be steeper compared with that of radio-quiet AGNs in previous works. In this work, we reinvestigate the fundamental plane in radio-loud AGNs by compiling a sample of 13 low-excitation radio galaxies (LERG) from the 3CR radio galaxies, for the reason that the accretion mode in LERG is believed to be a radiatively inefficient accretion flow. All the sources in our sample possess the data available at both the 5 GHz core radio luminosity detected  by VLA/VLBI/VLBA and the core X-ray luminosity detected by Chandra/XMM-Newton. Surprisingly, we find the slope in the fundamental plane ($\log L_{\rm R}=0.52 \log L_{\rm X}+ 0.84 \log M_{\rm BH} + 10.84$) of LERG is well consistent with that reported by \citet{m2003}. However, the normalization is found to be shifted by about 0.7 dex, which can be due to the difference on magnetic field strength in different objects. A shallower slope of $L_{\rm R}-L_{\rm X}$ relation ($L_{\rm R}\sim L_{\rm X}^{0.63}$) is also given by our sample, which demonstrates that the X-ray emission in LERG may come from accretion disc instead of jets as suggested by previous works.

\end{abstract}


\section{INTRODUCTION}

Accretion process is generally accepted to be the central engine of black hole astrophysical systems \citep{f2002}, e.g., black hole X-ray binaries (BHBs) and AGNs. In black hole systems with relativistic jets, a tight correlation between the X-ray and radio emissions ($L_{\rm R}\propto L_{\rm X}^{0.7}$) was reported by numerous works (e.g., \citealt{c2003,g2003,m2003,f2004}), where the X-ray and radio emissions are believed to come from accretion disc and jet, respectively. Coupled with black hole mass, a so-called fundamental plane (${\rm log}L_{\rm R}=0.6{\rm log}L_{\rm X}+0.78{\rm log}M_{\rm BH}+7.33$) was developed by \citet[][hereinafter, M03]{m2003} to manifest the activity of black hole and further explored by lots of following works (e.g., \citealt{f2004,m2006,l2008,d2015,n2016}).

While the fundamental plane is prevailing in recent years, there are still some noises remaining controversial. At first, the secular quasi-simultaneous observations on radio and X-ray fluxs in BHBs illustrated that their evolution can deviate obviously from the original M03 fundamental plane, which are known as 'outliers' \citep{x2007,c2011,c2013}. These outlier tracks possess a much steeper slope ($L_{\rm R}\propto L_{\rm X}^{1.4}$, see, e.g., \citealt{c2011}) compared with the original fundamental plane in radio-X-ray plane. The different radio-X-ray slopes may be originated from different accretion mode (the slopes of 0.7 and 1.4 correspond to radiatively inefficient and radiatively efficient accretion flows, respectively, \citealt{c2011, c2014}), or from the change of viscosity parameter $\alpha$ in a hot accretion flow \citep{x2016}. Secondly, it has long been suggested that the X-ray emission will be dominated by jet in the quiescent state of BHBs when the X-ray luminosity decrease to a critical value ($L_{\rm x,crit}/L_{\rm Edd}\sim 10^{-6}$, see \citealt{f2003,y2005}), resulting in a much steeper radio-X-ray slope (e.g., \citealt{w2007,p2013,r2014}). However, there are some works claiming that, even in quiescent state, the radio-X-ray correlation is still complied with the original M03 fundamental plane (\citealt{g2014,d2015}, but also see \citealt{x2017}). The last point, which is also the focus of this work, is the slope between radio and X-ray in radio-loud AGNs seems appear to be much steeper too compared with the radio-quiet AGNs \citep{w2006,l2008,d2011}, possibly due to the domination of strong jet emissions on radio and/or X-ray bands.

Radio galaxies (RG) can be divided into two classes according to the large-scale jet morphology traditionally, i.e., edge-darkened FRI and edge-brightened FRII \citep{f1974}. The difference between FRI and FRII can be originated from the interaction of jets with different power and their ambient mediums \citep{b1995,t2016}, and FRII usually possess higher jet power than FRI. Another important classification of RG is based on their optical spectroscopic information, where the RG with weak and strong emission lines are classified as LERG and high-excitation RG (HERG), respectively (e.g., \citealt{h1979,h2009}). HERG tend to have higher radio luminosity, similar with FRII.  However, there isn't an one-to-one match between these two classifications. Both FRI and FRII can comprise LERG and HERG (e.g., see \citealt{l1994}). From the observations of LERG, a radiatively inefficient accretion flow (RIAF) should be present due to their lack of AGN symbols, such as corona and torus, while HERG are believed to be powered by a radiatively efficient cold accretion disk \citep{h2007}. Similarly, low-luminosity AGNs are also found to be different in various aspects with bright AGNs (see, e.g., \citealt{h2008,g2009,s2011,x2011,l2017}). Therefore, we can naturally anticipate that the original M03 fundamental plane will change for HERG because of the transition of accretion mode. As a result, in order to investigate the radio-X-ray slope in radio-loud AGNs, we must discriminate LERG and HERG at first. We notice that in some previous works where a steeper slope was reported, the authors didn't distinguish LERG and HERG from radio-loud AGNs (e.g., \citealt{l2008}). In this work, we reinvestigate the fundamental plane of radio-loud AGNs by constructing a sample satisfying the following conditions: 1), since the accretion modes of LERG and HERG are different, all the sources in the sample should be LERG in order to ensure the accretion flow is RIAF. 2), radio emission mainly comes from jet in radio-loud AGNs, where both the core and lobe can play important roles. To avoid the influence of surrounding mediums, we adopt the source being core dominated only.

\section{The Sample of LERG}

RG can be divided into LERG and HERG according to their optical spectroscopic properties. \citet{l1994} suggested that the RG with $[O_{\rm III}]/H_{\rm \alpha}>0.2$ and an equivalent width of $[O_{\rm III}] > 0.3$ are HERG, while the LERG possess weak $[O_{\rm III}]$ lines. Following this advice, \citet{b2009,b2010} developed an excitation index (EI) as new spectroscopic indicator to discriminate LERG and HERG, where $\rm{EI = {log}(O_{III}/H_{\beta})- 1/3[log(N_{II}/H_{\alpha}) +}$ $\rm{log(S_{II}/H_{\alpha}) + log(O_{I}/H_{\alpha})] < 0.95} $ for LERG.

Our parent sample is the 113 3CR radio sources with redshift $z<0.3$, in which all the emission lines mentioned above are detected in 83 sources \citep{b2009}. We first exclude the 43 HERG with $\rm {EI > 0.95}$ because their accretion flows may be radiatively efficient, leading to 40 LERG. Radio flux in radio-loud AGNs is dominated by the synchrotron emission of jet based on the truncated disc-jet model (see \citealt{y2014} for a review), which has been successfully applied to the M03 fundamental plane. In order to prevent the contamination of lobe, only the sources with radio core emissions detected by VLBA/VLBE/VLA are included in this work. For X-ray, we adopt the sources with X-ray core flux detected by Chandra/XMM-Newton only to maintain the high precision. At last, we get 13 LERG with the core radio and X-ray emissions satisfied the requirement above (see table 1).

All the data of Cols (1), (2), (3), (4) and (8) are directly taken from \citet{h2016} except for 3C 442. We get the black hole mass of 3C 442 from \citet{d2015} and calculate its Eddington ratio of ionizing luminosity accordingly. The black hole mass of 3CR radio sources in \citet{h2016} is derived with the $M_{\rm bh}-L_{\rm bul}$ correlation \citep{marconi2003}, where $L_{\rm bul}$ is gotten from \citet{b2010}. Utilizing the data of emission lines from \citet{b2009}, the excitation index EI can also be estimated \citep{h2016}. We gathered the core radio luminosity $L_{\rm R}$ at 5 GHz in Col (5) from NED \footnote{http://ned.ipac.caltech.edu/}. For sources without direct observations at 5 GHz (labelled with an asterisk), the luminosity at 5 GHz is derived from the neighbouring frequencies with a spectral index $\alpha_{\rm r}\sim 0$ for low-luminosity AGNs (see, e.g., \citealt{u2001} and \citealt{h2008}). The nuclear X-ray luminosity $L_{\rm {x, 2-10 keV}}$ are also derived from NED, with photon index $\Gamma$ listed in table 1. We adopt the data from Chandra/XMM-Newton only to get the high precision.

\begin{table*}
\normalsize
\caption{The sample of LERG.}
\begin{minipage}{\textwidth}
\begin{center}
\begin{tabular}{llccllccc}
\hline
{Name} & {$z$} &  {EI} & {$\log M_{\rm bh}$} &  {$\log L_{\rm R}$} & {
$\Gamma$} & {$\log L_{\rm X}$}  & {$\log (L_{\rm ion}/L_{\rm Edd})$} & {Reference$^a$} \\
{(1)} &  {(2)} &  {(3)} &  {(4)} &  {(5)} &  {(6)} &  {(7)} &  {(8)} &  {(9)}\\
\hline

3C 31    &  0.017  &  0.797  & 9.1  &  39.41$^\ast$  & 1.4  & 40.74  & -4.88 & 2\\
3C 66B   &  0.017  &  0.746  & 9.4  &  39.90 & 2.4   & 41.03   & -4.63 & 2\\
3C 84    &  0.018  &  0.738  & 9.3  &  40.96 & 1.6   & 42.68   & -2.96 & 2\\
3C 88    &  0.03   &  0.588  & 8.7  &  40.19 & 1.11  & 41.22   & -3.87 & 5\\
3C 264   &  0.022  &  0.313  & 8.9  &  39.96 & 2.4   & 42.14   & -4.94 & 1\\
3C 270   &  0.007  &  0.048  & 8.8  &  39.33 & 1.09  & 41.22   & -5.15 & 1\\
3C 272.1 &  0.004  &  0.478  & 8.6  &  38.54 & 2.1   & 39.65   & -5.64 & 2\\
3C 274   &  0.004  &  0.233  & 9.0  &  39.96 & 2.3   & 40.64   & -5.24 & 2\\
3C 317   &  0.034  &  0.619  & 9.3  &  40.51 & 2.0   & 41.49   & -4.23 & 2\\
3C 338   &  0.032  &  0.288  & 9.4  &  40.01 & 2.15  & 42.37   & -5.09 & 1\\
3C 371   &  0.05   &  0.692  & 9.0  &  41.41 & 1.46  & 43.38   & -3.33 & 4\\
3C 442   &  0.026  &  0.713  & 8.4  &  39.41 & 1.4   & 40.51   & -4.47 & 3\\
3C 465   &  0.03   &  0.514  & 9.5  &  40.27$^\ast$  & 2.59  & 41.37  & -4.96 & 1\\

\hline
\end{tabular}
\end{center}
\end{minipage}
Notes: Col.(1): Source name. Col.(2): Redshift $z$. Col.(3): Excitation index EI.  Col.(4): Black hole mass $M_{\rm bh}$. Col.(5): Radio spectral luminosity at 5\ GHz, $L_{\rm R}$, in unit of ${\rm erg\ s^{-1}\ Hz^{-1}}$. Col.(6): Photon index $\Gamma$. Col.(7): X-ray spectral luminosity from 2-10 keV, $L_{\rm X}$, in unit of ${\rm erg\ s^{-1}\ Hz^{-1}}$. Col.(8): Eddington ratio of ionizing luminosity. \\
$^\ast$: For sources labelled with $^\ast$, their radio flux at 5 GHz are derived based on observations at neighbouring frequencies.\\
$^a$: The reference for photon index $\Gamma$ in our sample: (1) \citet{d2004}; (2) \citet{b2006}; (3) \citet{h2007}; (4) \citet{s2007}; (5) \citet{g2008}.
\end{table*}

\section{Fitting Method and Result}

\begin{figure}
\centering
\includegraphics[width=8.5cm]{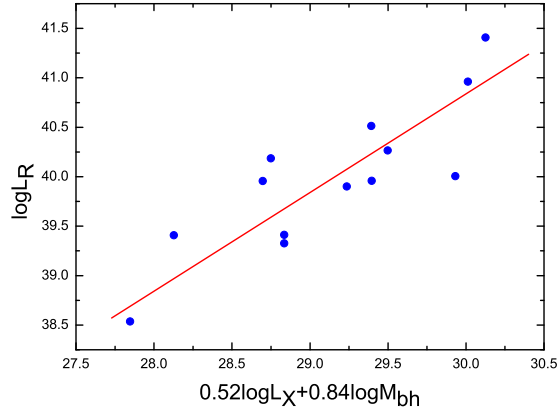}
\caption{The fundamental plane of black hole activity in LERG, where the solid line is the best fit. }\label{f1}
\end{figure}

\begin{figure}
\centering
\includegraphics[width=8.5cm]{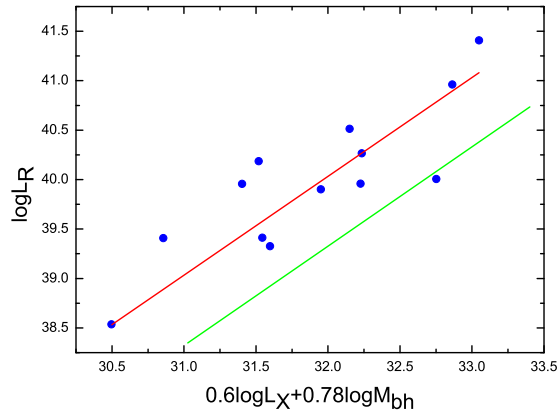}
\caption{The fundamental plane of black hole activity in LERG, where the green line shows the M03 fundamental plane relation. The red line indicates the movement of red line by 0.7 dex. }\label{f2}
\end{figure}

\begin{figure}
\centering
\includegraphics[width=8.5cm]{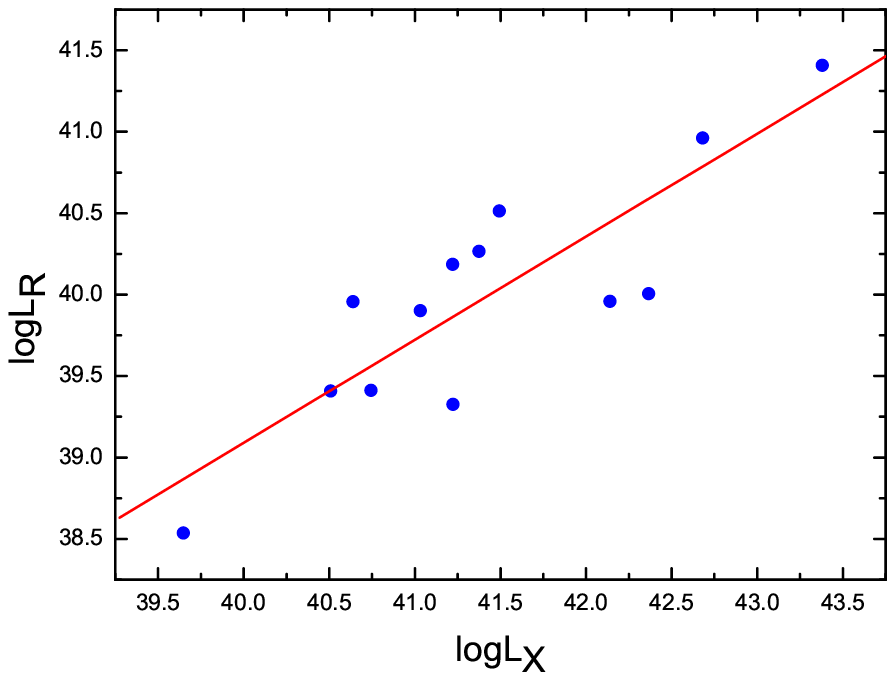}
\caption{The relationship between the radio luminosity $L_{\rm R}$ and X-ray luminosity $L_{\rm X}$ in LERG. }\label{f3}
\end{figure}

Following M03, we consider the linear relation below to reinvestigate the fundamental plane in radio-loud AGNs,
\begin{equation}
\log L_{\rm R}=\xi_{\rm X} \log L_{\rm X}+\xi_{\rm M}\log M_{\rm BH}+c,
\end{equation}
where $L_{\rm R}$ and $L_{\rm X}$ are the radio luminosity at 5 GHz and the 2-10 keV X-ray luminosity, respectively. A least $\chi^2$ method is adopted to fit the multivariate relation coefficients as in M03, which minimizes the following statistic,
\begin{equation}
\chi^2=\sum \frac{(\log L_{\rm R}-\xi_{\rm X} \log L_{\rm X}-\xi_{\rm M}\log M_{\rm BH}-c)^2}{\sigma_{\rm R}^2+\xi_{\rm X}^2\sigma_{\rm X}^2+\xi_{\rm M}^2\sigma_{\rm M}^2},
\end{equation}
where we adopt the uncertainties $\sigma_{\rm R}=0.2$, $\sigma_{\rm X}=0.3$, and $\sigma_{\rm M}=0.4$ according to the typical variations in AGN observations as in \citet{d2015}, instead of assuming isotropic uncertainties as in M03.

In figure \ref{f1}, we present the fundamental plane for our LERG sample through the least $\chi^2$ method. The best fitting result reads,
\begin{equation}
\log L_{\rm R}=0.52^{+0.16}_{-0.16} \log L_{\rm X}+ 0.84^{+0.50}_{-0.50} \log M_{\rm BH} + 10.84^{+5.95}_{-5.95},
\end{equation}
with an intrinsic scatter of $\sigma_{\rm int}=0.38$ dex. We find that the radio-X-ray slope of our sample is consistent with the original slope (0.6) in M03 quite well. In order to further verify the fundamental plane of our sample, we plot a figure to investigate their normalization as \citet{d2011} (Figure \ref{f2}). It is found that the normalization of our sample is indeed larger than that in M03 by about 0.7 dex, though this deviation is still within the range of their error bars. We suggest this movement can be due to the difference on magnetic field strength (see the last section for further discussion).

Furthermore, we investigate the relationship between $L_{R}$ and $L_{X}$ in our LERG sample (Figure \ref{f3}). A linear fit gives,
\begin{equation}
\log L_{\rm R}=(0.63\pm0.11)\log L_{\rm X}+13.78\pm4.71,
\end{equation}
with a strong confidence level larger than 99.9\% based on a Pearson test. This result is also consistent with previous works \citep{c2003,g2003,m2003,f2004}.

\section{Conclusion And Discussion}

In this work, we compile a sample of LERG from the 3CR radio galaxies with optical spectroscopic information \citep{b2009}. After excluding the sources with excitation index $\rm {EI} > 0.95$, a sample of 13 LERG is found to contain both the data of core radio luminosity at 5 GHz detected by VLA/VLBI/VLBA and core X-ray luminosity detected by Chandra/XMM-Newton. Surprisingly, We discover a similar radio-X-ray slope  with that of M03 fundamental plane, which suggests that the low-luminosity radio-loud AGNs (LERG) still follow the original M03 fundamental plane. We notice that \citet{d2011} had investigate the fundamental plane in a LERG sample either. They discovered a steeper radio-X-ray slope and advised the X-ray emissions in LERG may be originated from jets, though their X-ray luminosity $L_{\rm X}$ are larger than $L_{\rm crit}$ (see below). The reason for this inconformity may be that their sample also comprised some steep spectrum LERG except for the core dominated flat spectrum LERG.

Furthermore, we find the normalization of our sample is larger than that in M03 by about 0.7 dex, though this deviation is still within the range of their error bars. The possible reason for this movement can be the variant magnetic field strength in different objects. If we consider the parameter $\beta$ (the ratio of gas pressure to magnetic pressure) isn't a constant in different objects, the radio flux from jet can be roughly written as $L_{\rm R}\propto \dot{m}^{1.4} \beta^{-1}$ ($L_{\rm R}\propto \dot{m}^{1.4}$ when $\beta$ is constant, see \citealt{h2003}). The X-ray flux can be revised as $L_{\rm X}\propto \dot{m}^{2} \beta^{a}$ ($a>0$) for the same way, because the X-ray flux increase when the magnetic field strength decrease (see \citealt{m1997}). Therefore, the revised $L_{\rm R}$-$L_{\rm X}$ relation can be given as:
\begin{equation}
\log L_{\rm R}\propto 0.7 \log L_{X}-(1+0.7a)\log \beta.
\label{mag}
\end{equation}
All the objects in our sample are radio-loud, which means higher magnetic field strength and then smaller $\beta$. According to equation (\ref{mag}), we can naturally anticipate a higher normalization for radio-loud objects. Indeed, the large-scale magnetic field is easy to be magnified in a RIAF due to their high radial velocity \citep{l1994,c2011b,l2014} and can strongly affect the activity of black hole. Except for the obvious augment in radio emission, the X-ray emission in a RIAF is a complicated function of the magnetic field strength based on the theoretical research \citep{n1994,n1995,m1997,b2013,b2016,s2015}. Furthermore, large-scale magnetic field can change the value of viscosity parameter $\alpha$ according to the recent magneto hydrodynamical (MHD) simulations \citep{bs2013,s2016}, which can further decrease the X-ray emission of RIAF (e.g., \citealt{n1994,l2017}). These points will be explored in future works.

The slope of $L_{\rm R}-L_{\rm X}$ correlation in LERG is also found to be consistent with other black hole systems (e.g., \citealt{c2003,g2003,m2003,f2004}), but much shallower than that found in FRI samples (e.g., \citealt{d2015}). In theory, \citet{y2005} suggested that there is a critical X-ray luminosity ($L_{\rm crit} = L_{\rm {X,2-10 keV}} \sim 10^{-6} L_{\rm Edd}$) to diagnose the origin of X-ray in low-luminosity AGNs. When $L_{\rm X} > L_{\rm crit}$, the X-ray from accretion disc will exceed that from jet, resulting on a shallower slope between the relation of $L_{\rm X}$ and $L_{\rm R}$. The Eddington ratio of ionizing luminosity $L_{\rm ion}/L_{\rm Edd}$ in our sample, which is a significant fraction of the bolometric luminosity \citep{w1999}, is all larger than $L_{\rm crit}$. Therefore, the shallower slope of $L_{\rm X}-L_{\rm R}$ correlation in Fig. \ref{f3} demonstrates that the X-ray emission should come from a RIAF in LERG. Our results indicate that, considering the core emissions of radio and X-ray, the radio-loud AGNs still comply with the physics of truncated accretion disc-jet model (e.g., \citealt{y2014}), which had been successfully applied in low-luminosity AGNs.

\section*{Acknowledgements}

We thank the referee Dr. A. Merloni for helpful comments and suggestion. SLL thanks Dr. Fu-Guo Xie for useful discussion and providing the code for the least $\chi^2$ method. This work is supported by the Natural Science Foundation of China (grants 11773056, 11473054 and U1531245), the Youth Innovation Promotion Association of the Chinese Academy of Sciences (CAS) (ids. 2015216) and the Key Research Program of Frontier Sciences of CAS (No. QYZDJ-SSW-SYS023). This work has made extensive use of the NASA/IPAC Extragalactic Database (NED), which is operated by the Jet Propulsion Laboratory, California Institute of Technology, under contract with the National Aeronautics and Space Administration (NASA).

\label{lastpage}

\end{document}